\documentclass[12pt,preprint]{aastex}
\usepackage{epsfig}
\newcommand       \Angstrom     {\,{\rm \AA}} 
          
\newcommand       \cm           {\,{\rm cm}}

\newcommand	  \g		{\,{\rm g}}

\newcommand	  \yr		{\,{\rm yr}}

\newcommand       \simlt        {\lesssim}

\newcommand       \mum          {\,{\rm \mu m}}
\newcommand	  \ppm		{\,{\rm ppm}}

\newcommand	  \NH		{N_{\rm H}}
\newcommand       \simali       {\sim\,}

\newcommand	  \tauext	{\tau_{\rm ext}}
\newcommand	  \Vtot	        {V_{\rm dust}^{\rm tot}}

\newcommand	  \magni	{\,{\rm mag}}
\newcommand	  \eeff	        {\epsilon_0^{\rm eff}}

\newcommand	  \xism         {\left[{\rm X/H}\right]_{\rm ISM}}
\newcommand	  \csun         {\left[{\rm C/H}\right]_{\odot}}

\newcommand	  \osun         {\left[{\rm O/H}\right]_{\odot}}

\newcommand	  \cism         {\left[{\rm C/H}\right]_{\rm ISM}}

\newcommand	  \xdust        {\left[{\rm X/H}\right]_{\rm dust}}
\newcommand	  \cdust        {\left[{\rm C/H}\right]_{\rm dust}}

\newcommand	  \fedust       {\left[{\rm Fe/H}\right]_{\rm dust}}
\newcommand	  \mgdust       {\left[{\rm Mg/H}\right]_{\rm dust}}
\newcommand	  \sidust       {\left[{\rm Si/H}\right]_{\rm dust}}
\newcommand	  \xgas         {\left[{\rm X/H}\right]_{\rm gas}}
\newcommand	  \cgas        {\left[{\rm C/H}\right]_{\rm gas}}

\newcommand	  \mux         {\mu_{\rm X}}

\newcommand{\figwidth}{4.0in}

\countdef\decade=200
\advance\decade by \year
\countdef\hours=201
\advance\hours by \time
\divide\hours by 60
\countdef\mins=202
\advance\mins by \hours
\multiply\mins by 60
\multiply\hours by 100
\countdef\miltime=203
\advance\miltime by \hours
\advance\miltime by \time
\advance\miltime by -\mins

\shorttitle{Fluffy Dust versus Subsolar Interstellar Abundances}

\begin{document}

\title{
{\normalsize\rm To appear in {\it The Astrophysical Journal},
  vol.\,622 (the April-01-2005 issue)}\\
 \vspace*{1.0em}
Can Fluffy Dust Alleviate
the ``Subsolar'' Interstellar Abundance Problem?
	 }
\author{Aigen Li\altaffilmark{1,2}}
\altaffiltext{1}{Theoretical Astrophysics Program,
                 Lunar and Planetary Laboratory 
                 and Steward Observatory,
                 University of Arizona, Tucson, AZ 85721;
                 {\sf agli@lpl.arizona.edu}}
\altaffiltext{2}{Department of Physics and Astronomy,
                 University of Missouri, Columbia, MO 65211;
                 {\sf LiA@missouri.edu}}

\begin{abstract}
What might be the most appropriate set of interstellar 
reference abundances of chemical elements 
(both in gas and in dust) relative to hydrogen has been 
a subject of much discussion in the past decade. 
While historically the Sun has been 
taken as the reference standard,
it has recently been suggested that the interstellar 
abundances might be better represented by those of
B stars (because of their young ages)
which are just $\simali$60--70\% of the widely
adopted solar values (``subsolar'').
On the other hand, the most recent estimates 
of the solar carbon and oxygen abundances are 
also close to those of B stars.
If the interstellar abundances are indeed ``subsolar''
like B stars or the newly-determined solar C,O values, 
there might be a lack of raw material to form 
the dust to account for the interstellar extinction. 
In literature it has been argued that 
this problem could be solved if interstellar
grains have a fluffy, porous structure
since fluffy grains are more effective in 
absorbing and scattering optical and ultraviolet 
starlight than compact grains (on a per unit mass basis). 
However, we show in this work that, 
using the Kramers-Kronig relation, 
{\it fluffy dust is not able to overcome 
the abundance shortage problem}.
A likely solution is that the abundances of refractory 
elements in {\it stellar photospheres}
may under-represent the composition of the interstellar 
material from which stars are formed, 
resulting either from the possible underestimation of
the degree of heavy-element settling in stellar
atmospheres, or from the incomplete incorporation
of heavy elements in stars during the star formation
process.
\end{abstract}
\keywords{dust, extinction --- ISM: abundances}

\section{Introduction\label{sec:intro}}
Elements in the interstellar medium (ISM) 
generally exist in the form of gas or dust.
The interstellar gas-phase abundances of elements
can be measured from their optical and ultraviolet (UV)
spectroscopic absorption lines. The elements ``missing'' 
from the gas phase are bound up in dust grains --
this phenomenon is often called ``interstellar depletion''. 
The determination of the dust-phase abundances (``depletion'') 
is indirect and more complicated. One usually relies on
the interstellar extinction modeling or the solid-state
spectral feature analysis. Both methods require 
an explicit assumption of the nature (e.g. grain 
composition, size and geometry) and the physical 
characteristics of the grains (e.g. their optical properties 
and the strengths of their characteristic vibrational bands
usually determined from their laboratory analogs). 
More commonly, the dust-phase abundance of an element is 
derived by assuming a reference abundance 
-- total abundance of this element 
(both in gas and in dust) for the ISM\footnote{%
   In astrophysical literature, the total abundance of elements 
   are also known as ``standard abundances'', ``interstellar 
   abundances'',  and ``cosmic abundances''.
   } 
-- and then from which subtracting off 
the observed gas-phase abundance. 
 
Interstellar depletions allow us to extract important 
information about the composition and quantity of interstellar 
dust (see \S2.11 in Li 2004a):
(1) The fact that Si, Fe, Mg, C, and O are depleted
in low density clouds indicates that interstellar dust 
must contain an appreciable amount of Si, Fe, Mg, C and O. 
Indeed, all contemporary interstellar dust models consist 
of both silicates and carbonaceous dust (see Li 2004b for
a review). 
(2) From the depletion of the major elements Si, Fe, Mg, C, 
and O one can estimate the gas-to-dust mass ratio to be 
$\simali$165 (see Li 2004a), if the interstellar abundances 
are assumed to be those of the solar values of Holweger (2001). 
(3) In addition to the silicate dust component, there must 
exist another dust population, since silicates alone are 
not able to account for the observed amount of extinction 
relative to H although Si, Mg, and Fe are highly depleted 
in the ISM. Even if all Si, Fe, and Mg elements 
of solar abundances of Holweger (2001) 
are locked up in submicron-sized silicate grains, 
they can only account for $\simali$60\% of 
the total observed optical extinction (see Li 2004a).

Apparently, in interstellar depletion and dust composition 
studies the knowledge of interstellar reference abundances 
is critical. Historically, the solar abundances have been
taken to represent the total interstellar abundances.
But we note that the published solar abundances have undergone 
major changes over the years and are still subject to major 
systematic uncertainties. This is demonstrated in 
Table \ref{tab:sunabd} in which the widely used 
solar abundances over the past 3 decades for 
the dust-forming elements C, O, Mg, Si and Fe are tabulated. 
Most notably, the most recent estimates of the solar
C ($\csun \approx 245\ppm$; Allende Prieto, Lambert,
\& Asplund 2002) and O abundances ($\osun \approx 457\ppm$;
Asplund et al.\ 2004) are significantly reduced
from their earlier values.

Using the gas-phase abundances measured by 
the {\it Copernicus} satellite for the $\zeta$ Ophiuchi 
sightline  (Morton et al.\ 1973) and the solar abundances 
of Cameron (1973; see Table \ref{tab:sunabd})
as the reference abundances,
Greenberg (1974) found that the observed depletion 
of C, N, and O is significantly greater than 
could be accommodated by the dust under 
any reasonable models.\footnote{%
  Whittet (1984) argued that the {\it Copernicus}
  determinations of the C, N, and O depletions 
  for $\zeta$ Oph were subject to systematic errors 
  arising in the gas-phase abundance analysis 
  and the depletions toward many other stars 
  are actually substantially lower.
  }
Twenty years later, Sofia, Cardelli, \& Savage (1994) 
found that the interstellar depletions are lowered 
for C, N, and O 
if B stars are used as the reference standard.
They argued that the solar system may have 
enhanced abundances of many elements, 
and therefore the solar abundances
(of Anders \& Grevesse 1989) 
are not representative of the interstellar abundances.

Snow \& Witt (1996) analyzed the surface abundances of
B stars and field F and G stars and found that not only
C, N, and O but also Si, Mg, and Fe and many other elements
are underabundant in these stars. This led them to suggest
that the interstellar abundances are appreciably 
subsolar ($\simali$60\%--70\% of the solar values compiled
by Anders \& Grevesse 1989). It is worth noting that the most 
recent estimates of the solar C ($\csun \approx 245\ppm$; 
Allende Prieto et al.\ 2002) and O abundances 
($\osun \approx 457\ppm$; Asplund et al.\ 2004)\footnote{%
  The substantial drop in $\osun$ is of particular interest:
  the original $\simali$851$\ppm$ (Anders \& Grevesse 1989) 
  was considered too large to be accounted for in the ISM
  if the solar abundance is taken to be the interstellar 
  reference standard (Mathis 1996). This was one of the major 
  motivations to invoke subsolar interstellar abundances.
  }
are also ``subsolar'', just $\simali$50\%--70\% of 
the Anders \& Grevesse (1989) values
and close to the ``subsolar'' interstellar abundances
originally recommended by Snow \& Witt (1996). 
These reductions in the amounts of heavy elements available 
for grains have turned the former surplus of raw materials 
(Greenberg 1974) into a shortage and therefore have profound 
significance for the nature of interstellar dust (Snow \& Witt 1996). 

If the interstellar abundances are indeed ``subsolar''
like B stars or the newly-determined solar values, 
there is clearly a crisis for interstellar 
grain modeling; in particular, this places severe challenges to 
the interstellar extinction modeling\footnote{%
  Most grain models require more elements to be tied up in 
  dust than available if the interstellar abundances are 
  indeed significantly subsolar (see Snow \& Witt 1995, 1996).
  }
and the interpretation of the interstellar 9.7$\mum$ 
feature which is generally attributed to amorphous 
silicate dust\footnote{%
 The amount of Si (relative to H) required to deplete in 
 dust to account for the observed 9.7$\mum$ feature strength 
 $\Delta \tau_{9.7\mum}/A_V$$\approx 1/18.5$ 
 in the local diffuse ISM is
 \begin{equation}
 \left[\frac{\rm Si}{\rm H}\right] = 
 \frac{\Delta \tau_{9.7\mum}}{N_{\rm H}}
 \frac{1}{\kappa_{\rm sil}^{\rm abs}(9.7\mum)\,\mu_{\rm sil}}
 = \frac{\Delta \tau_{9.7\mum}}{A_V}
   \frac{A_V}{N_{\rm H}} 
   \frac{1}{\kappa_{\rm sil}^{\rm abs}(9.7\mum)\,\mu_{\rm sil}}
 \end{equation}
 where $\kappa_{\rm sil}^{\rm abs}(9.7\mum)$ is the silicate
 mass absorption coefficient at $\lambda$=9.7$\mum$,
 $A_V$ is the visual extinction, and $N_{\rm H}$ is the hydrogen 
 column density, $\mu_{\rm sil}$ is the silicate molecular weight.
 With $\kappa_{\rm sil}^{\rm abs}(9.7\mum)$$\approx$2850$\cm^2\g^{-1}$
 and $\mu_{\rm sil}$$\approx$172$\mu_{\rm H}$ for amorphous
 olivine MgFeSiO$_4$, the local diffuse ISM
 ($A_V/N_{\rm H}$$\approx$$5.3\times$$10^{-22}\magni\cm^2$)
 requires ${\rm Si}/{\rm H}$$\approx$35$\ppm$,
 significantly exceeding the B star Si/H abundance
 ($\approx 18.8\ppm$; Sofia \& Meyer 2001).
 But also see Mathis (1998) who argued that composite
 fluffy spheroids with axis ratios greater than 2 
 could explain the observed silicate features even
 if the interstellar abundances are subsolar. 
 }
-- this is often referred to as interstellar abundance 
``budget crisis'' (e.g. see Kim \& Martin 1996).

In the context of these reduced reference abundances, 
Mathis (1986) re-investigated the composite interstellar 
dust model and found that fluffy grains consisting
of small silicates, oxides, amorphous carbon derived from
burning benzene (``BE''), and vacuum with a volume fraction 
of $\simali$45\% could account for the observed interstellar 
extinction and satisfy the much tighter abundance constraints.
Other dust models invoking compact grains require more 
elements than available (Li \& Greenberg 1997, Weingartner 
\& Draine 2001, Li \& Draine 2001, Zubko, Dwek, \& Arendt 2004).
By comparing the observed dust-to-gas ratio of the $\zeta$ Oph
sightline with that expected from the reduced depletions,
Snow \& Witt (1996) concluded that the dust in this sightline
must have a fluffy, porous or fractal structure.  

However, can fluffy dust really alleviate the subsolar
interstellar abundance budget crisis? We will explore
this issue in terms of the Kramers-Kronig relations
(Purcell 1969), using the most recently determined
(subsolar) B star abundances (Sofia \& Meyer 2001; 
see Table \ref{tab:sunabd}) 
as the reference standard.\footnote{%
  Although the most recently determined solar C and O 
  abundances are in good agreement with those of B stars,
  and both are substantially lower than the widely
  adopted Anders \& Grevesse (1989) solar values,
  for simplicity we customarily call the B star
  and new solar (C, O) abundances ``subsolar''.  
  }
It is found in this paper
that the Kramers-Kronig relations readily rule out 
the B star abundances as the interstellar reference 
abundances since the available elements are insufficient 
to account for the observed interstellar extinction, 
independent of specific grain models and grain fluffiness. 
It is also found that if the Sun (of Holweger 2001)
or the field F and G stars (of Sofia \& Meyer 2001)
are used as the reference standard, 
the dust depletion is not inconsistent 
with the interstellar extinction. 
However, given that even the elemental abundances 
of our own Sun are not well determined,
we are not at a position to conclude that
the Sun or the field F and G stars reference abundances 
are to be preferred; after all, if the solar Mg, Si,
and Fe elements are also close to these of B stars 
as the newly determined solar C and O abundances, 
the solar abundances would also fall short in 
accounting for the interstellar extinction.  

\section{Constraints from the Kramers-Kronig Relations\label{sec:kk}}
As shown by Purcell (1969), the Kramers-Kronig relations 
can be applied to the interstellar space sparsely populated 
by interstellar grains. 
Let $\tauext(\lambda)$ be the extinction optical depth
at wavelength $\lambda$ 
and $\tauext(\lambda)/\NH$ be the extinction per H nucleon.
The Kramers-Kronig relations can be used to 
relate $\int_{0}^{\infty} \tauext(\lambda)/\NH\,d\lambda$
to the total volume occupied by dust per H nucleon 
$\Vtot/{\rm H}$ through 
\begin{equation}\label{eq:kk}
\int_{0}^{\infty} \frac{\tauext(\lambda)}{\NH} d\lambda 
= 3 \pi^2 F \frac{\Vtot}{\rm H} ~~,
\end{equation}
where the dimensionless factor $F$ is the orientationally-averaged 
polarizability relative to the polarizability of an equal-volume 
conducting sphere, depending only upon the grain shape and 
the static (zero-frequency) dielectric constant 
$\epsilon_0$ of the grain material 
(Purcell 1969; Draine 2003a).

Let $P$ be the volume fraction of vacuum (``fluffiness''
or ``porosity'') contained in a fluffy grain. 
Let $\xism$ be the total interstellar abundance 
of element X relative to H, $\xgas$ be the amount 
of X in gas phase, $\xdust$ be the amount of X 
contained in dust (obviously we have $\xdust$\,=\,$\xism-\xgas$),
and $\mux$ be the atomic weight of X. 
Let there be $N$ dust species in a composite fluffy grain;
let $\rho_j$ be the mass density of dust species $j$; 
let $f_{{\rm X},j}$ be the fraction of element X
locked up in dust species $j$. 
For a chosen set of interstellar reference abundances,
the total volume of interstellar dust per H nucleon can 
be estimated from the interstellar depletions
\begin{equation}\label{eq:vtot}
\frac{\Vtot}{\rm H} = \sum_{j=1}^{N}\sum_{\rm X} 
f_{{\rm X},j} \left(\xism-\xgas\right) 
\mu_{\rm X}/\left(1-P\right) \rho_j 
\end{equation}
where the first summation is over all possible dust species
and the second one is over all condensable elements.\footnote{%
  Eq.(\ref{eq:vtot}) actually slightly overestimates the total
  dust volume since the $1/(1-P)$ enlarging factor has also
  been applied to the ultrasmall grains [e.g. polycyclic aromatic
  hydrocarbon molecules (PAHs) responsible for the 3.3, 6.2, 7.7, 8.6
  and 11.3$\mum$ ``unidentified infrared'' (UIR) emission features],
  which are separated from the bulk, composite fluffy grains
  which contain most of the mass.
  Although small PAHs are expected to be planar 
  (and thus ``highly flattened''), their static
  dielectric constants $\epsilon_0$ are not large 
  (e.g. $\epsilon_0\approx 2.3$ for benzene),
  therefore the $F$ factors for PAHs will be 
  in the order of unity and their contribution
  to the right-hand-side of Eq.(\ref{eq:kk}) 
  will not be large, after all, in the silicate-graphite-PAHs
  interstellar grain model (Weingartner \& Draine 2001;
  Li \& Draine 2001) the PAH population takes over
  only $\approx 6.1\%$ of the total grain volume
  (and $\approx 4.6\%$ of the total grain mass).  
  }

The interstellar extinction per H nucleon $\tauext(\lambda)/\NH$
is known for a limited range of wavelengths. 
For the diffuse ISM, the mean extinction per H nucleon
$\tauext(\lambda)/\NH$ is fairly well-determined from 
the far UV to infrared (IR): 
$0.1\mum \simlt \lambda \simlt 30\mum$ (Draine 2003b). 
For $30\mum \simlt \lambda \simlt 1000\mum$
we will adopt the theoretical $\tauext(\lambda)/\NH$ values
calculated from the silicate-graphite-PAH model which has
been shown to successfully reproduce the observed interstellar 
extinction from the far-UV to mid-IR and the observed IR emission
(Weingartner \& Draine 2001, Li \& Draine 2001). 
Since $\tauext(\lambda)$ is a positive number for all
wavelengths, the integration of $\tauext(\lambda)/\NH$
over a finite wavelength range can be used to obtain
a lower bound on $F/\left(1-P\right)$
\begin{equation}\label{eq:FP}
\left(\frac{F}{1-P}\right)_{\rm min} = \frac{
\int_{912\Angstrom}^{1000\mum} \tauext(\lambda)/\NH\,d\lambda} 
{3 \pi^2 \sum_{j=1}^{N}\sum_{\rm X} 
f_{{\rm X},j} \left(\xism-\xgas\right) 
\mu_{\rm X}/\rho_j} ~~.
\end{equation}

At a first glance of Eqs.(\ref{eq:kk},\ref{eq:vtot}),
there appears to be no problem for dust models with subsolar
interstellar abundances to account for the interstellar 
extinction, provided that the dust is sufficiently fluffy
(i.e., the porosity $P$ is sufficiently large). 
However, one should keep in mind that for a composite fluffy
grain, the increase in the total dust volume $\Vtot$
will be offset by a decrease in $F$ since the effective static 
dielectric constant becomes smaller when the dust becomes 
more porous which leads to a smaller $F$ (see Fig.\,1
in Purcell [1969] and Fig.\,15 in Draine [2003a]).

For the diffuse ISM, the integration of the mean
extinction over the wavelength range 
$912\Angstrom \le \lambda \le 1000\mum$ is approximately 
$\int_{912\Angstrom}^{1000\mum} \tauext(\lambda)/\NH\,d\lambda
\approx 1.37\times 10^{-25}\cm^{3}/{\rm H}$.
If we adopt the B star abundances as the reference
standard and subtract the observed gas-phase abundances
(see Table 2 in Sofia 2004), we obtain
$\cdust = \cism-\cgas\approx 60$ parts per million (ppm),
$\mgdust \approx 21\ppm$,
$\sidust \approx 16.8\ppm$,
and $\fedust \approx 27.5\ppm$.
It is reasonable to assume that all the dust-phase Si atoms
are incorporated into amorphous olivine silicates
with a stoichiometric composition of MgFeSiO$_4$.
This also consumes 16.8$\ppm$ Mg and Fe.   
We take the remaining 4.2$\ppm$ Mg to be depleted
in MgO, and the remaining 10.7$\ppm$ Fe evenly tied
up in FeO, Fe$_2$O$_3$, and Fe$_3$O$_4$. 
We will assume the 60$\ppm$ dust-phase carbon 
to be bound up either in graphite or in amorphous 
carbon. Therefore, from Eq.(\ref{eq:FP}) we obtain
$\left(F/\left[1-P\right]\right)_{\rm min}\approx 2.07$
if the carbon is in the form of graphite,
or $\left(F/\left[1-P\right]\right)_{\rm min}\approx 1.94$
if the carbon is in the form of ``BE'' amorphous carbon. 
In doing so, we take the mass density of
amorphous olivine MgFeSiO$_4$, MgO,
FeO, Fe$_2$O$_3$, Fe$_3$O$_4$, graphite, and
amorphous carbon to be 3.5, 3.58, 5.7, 5.25, 5.18,
2.24, and 1.8$\g\cm^{-3}$, respectively.

For a composite fluffy grain consisting of
small amorphous silicates, oxides (MgO, FeO, Fe$_2$O$_3$, 
and Fe$_3$O$_4$), and graphite or amorphous carbon particles,
we use the Bruggeman effective medium theory 
(Bohren \& Huffman 1983; Ossenkopf 1991) to calculate 
its effective static dielectric constants $\eeff$.
The static dielectric constants of the constituent
dust materials of the composite grain adopted in this
work are: $\epsilon_0\approx 10$ for amorphous olivine
(Draine \& Lee 1984), $\epsilon_0\approx 9$ for MgO
(Roessler \& Huffman 1998), $\epsilon_0\approx 16$ for 
Fe$_2$O$_3$ (Steyer 1974), $\epsilon_0\approx 160$ for 
amorphous carbon (Rouleau \& Martin 1991), 
and $\epsilon_0\rightarrow \infty$ for FeO,
Fe$_3$O$_4$ and graphite (in the calculation, 
we take $\epsilon_0=10^{300}$ to represent
$\epsilon_0\rightarrow \infty$; it is found that 
there is not much difference among the model results 
obtained using $\epsilon_0=10^{100}$, 
$\epsilon_0=10^{200}$, or $\epsilon_0=10^{300}$).

Interstellar grains are nonspherical as indicated by
the detection of interstellar polarization. For simplicity,
we approximate these nonspherical grains by prolates or 
oblates. Interstellar polarization modeling suggests
that interstellar grains are modestly elongated:
Lee \& Draine (1985) found that the 3.1$\mum$ ice 
polarization feature of the Becklin-Neugebauer (BN) object
is best fit by $a/b=1/2$ oblates, where $a$ ($b$) is
the semiaxis along (perpendicular to) the symmetry axis;
similarly, Hildebrand \& Dragovan (1995) also found that 
$a/b=1/2$ oblates provide the best match to the 9.7$\mum$ 
silicate polarization feature of the BN object; 
using ice-coated core-mantle grains, Greenberg \& Li (1996) 
found that $a/b=3$ prolates are preferred in explaining the
9.7 and 18$\mum$ silicate polarization features of this object.
 
\begin{figure}[ht]
\begin{center}
\epsfig{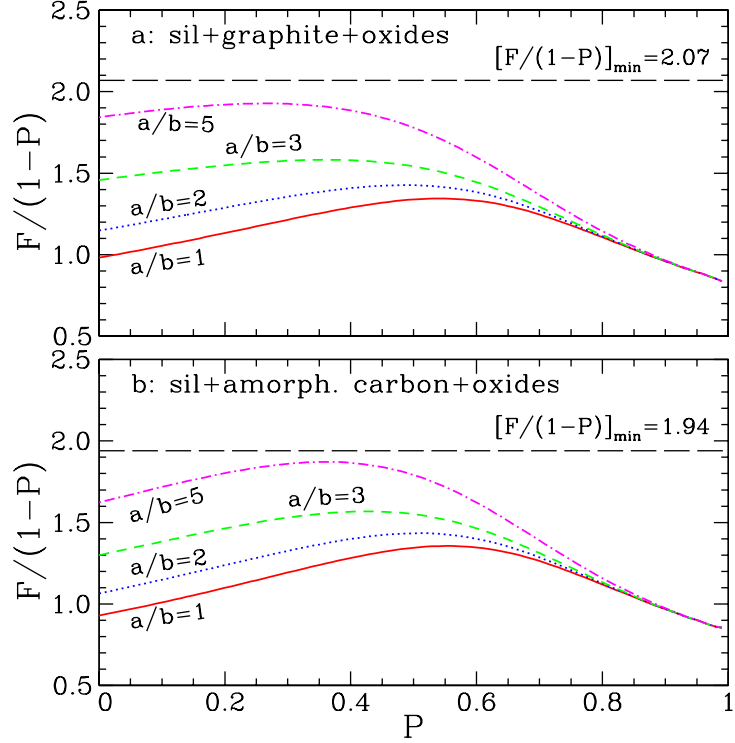}
\end{center}\vspace*{-1em}
\caption{
        \label{fig:FP}
        \footnotesize
        Model-predicted $\left(F/\left[1-P\right]\right)_{\rm mod}$
        as a function of porosity $P$
        for spherical (solid lines) or prolate composite fluffy 
        grains consisting of 
        (a) small silicates, oxides, and graphite; 
        or (b) small silicates, oxides, and amorphous carbon 
        with an elongation of $a/b=2$ (dotted lines),
        $a/b=3$ (dashed lines), and $a/b=5$ (dot-dashed lines),
        using the B star abundances as the interstellar
        reference standard.
        The horizontal long-dashed lines plot 
        $\left(F/\left[1-P\right]\right)_{\rm min}$,
        the lower limit on $\left(F/\left[1-P\right]\right)$
        required by the interstellar extinction
        (see Eq.[\ref{eq:FP}]).
        }
\end{figure}

For a composite fluffy grain of a given porosity $P$ and
a given elongation $a/b$, we first calculate its effective
static dielectric constant $\eeff$ and then calculate the
$F$ factor. In Figure \ref{fig:FP} we show the model-predicted
$\left(F/\left[1-P\right]\right)_{\rm mod}$ as a function 
of $P$ obtained for spheres ($a/b=1$) and prolates ($a/b=2,3,5$)
using the B star abundances as the interstellar reference standard. 
It is seen that even for $a/b=5$, the model values  
$\left(F/\left[1-P\right]\right)_{\rm mod}$ are always below
the lower limit $\left(F/\left[1-P\right]\right)_{\rm min}$,
no matter what chemical form the carbon takes
(graphite or amorphous carbon).
Similar results are obtained for oblate grains.
This indicates that {\it if the total interstellar abundances are
those of the B stars, there are not enough raw materials
to make the dust producing the interstellar extinction},
unless the grains are highly elongated or flattened
($a/b>6.1$ for prolates or $b/a>5.7$ for oblates).

It is also seen in Figure \ref{fig:FP} that 
the model value of $F/\left(1-P\right)$ first
increases with $P$ and reaches its maximum at
$P\approx 0.35$--0.55, depending on the grain shape,
and then decreases. This explains why in the latest 
version of the composite dust model -- to make the most 
economical use of the heavy elements -- Mathis (1996) arrived 
at a porosity of $P\approx 0.45$, in contrast to the original 
value of $P\approx 0.8$ (Mathis \& Whiffen 1989).\footnote{%
 The reason why our model-predicted $F/\left(1-P\right)$ 
 peaks at $P\approx 0.55$ for spherical grains consisting
 silicates, oxides and amorphous carbon (see Fig.\,\ref{fig:FP})
 while it is at $P\approx 0.45$ in the Mathis (1996) model
 lies in the fact the B star abundances and the interstellar
 gas-phase abundances we are using are somewhat different  
 from those adopted in Mathis (1996).
 }
For $P>0.8$, the model-predicted values of $F/\left(1-P\right)$ 
are insensitive to grain shape. This is because for grains with
a porosity $P>0.8$, their effective static dielectric constants
becomes smaller than $\simali$1.8, while the $F$ factor is 
insensitive to grain shape when $\epsilon_0<2$ 
(see Fig.\,1 of Purcell [1969] and Fig.\,15 of Draine [2003a]).

\section{Discussion\label{sec:discussion}}
Using the Kramers-Kronig relations, we have shown in 
\S\ref{sec:kk} that the B star abundances as the interstellar
reference standard fall short in accounting for 
the interstellar extinction, no matter how fluffy the dust
is or what precise chemical form the dust takes, 
unless the dust is highly elongated or flattened. 
Mathis (1996) suggested that the interstellar extinction
could be accounted for if the dust has a fluffy, porous
structure, consisting of small silicates, amorphous carbon, 
and oxides with vacuum comprising $\simali$45\% of its
volume. But Dwek (1997) argued against this suggestion
on the basis of the fluffy dust model emitting too much 
in the far-IR to be consistent with that observed for 
the diffuse ISM and the neglect of PAHs in the model 
(which are required to explain the ``UIR'' emission bands).\footnote{%
  Jones (1988) found that although the introduction of 
  a porous structure to a grain does increase its UV and 
  and far-UV absorption (per unit dust mass) compared to 
  its compact counterpart, its long wavelength absorptivity 
  is reduced. This is expected from the Kramers-Kronig relation 
  which indicates that the integration of the extinction cross 
  sections over the entire wavelength range should be proportional 
  to the total dust volume $V$ times the $F$ factor, unless the dust 
  is {\it both} extremely elongated {\it and} conducting, 
  $VF$ for porous grains does not differ much from that for 
  compact grains of the same mass. The enhanced short-wavelength 
  absorptivity of porous dust revealed in Jones (1988) also explains 
  why porous dust emits more in the far-IR: the dust absorbs more at 
  short wavelengths, and the absorbed energy must be radiated away 
  at long wavelengths!
  }
While the Dwek (1997) argument involves estimating 
the relatively uncertain UV and optical interstellar 
radiation field times the otherwise unobserved absorption 
(in contrast to extinction) of the grains (cf. Mathis 1998), 
the discussion presented in this work is more robust 
since essentially no assumptions are made 
in using the Kramers-Kronig relations to rule out 
the B star abundances as the interstellar reference standard.
The only assumption made in our analysis is that all the dust-phase 
Si atoms, with a similar number of Fe and Mg atoms,
are locked up in amorphous olivine silicates (MgFeSiO$_4$)
and the remaining Fe and Mg atoms form oxides. 
But we have also considered silicates in the form of 
amorphous pyroxene (Mg$_{\rm 1-x}$Fe$_{\rm x}$SiO$_3$) 
and amorphous olivine [Mg$_{\rm 2(1-x)}$Fe$_{\rm 2x}$SiO$_4$]  
with various Fe fraction ${\rm 0\le x\le 1}$ and even with 
some of the Si atoms in SiO$_2$. It is found that the conclusion 
of this work remains unchanged, unless an appreciable
fraction of the Fe atoms are in the form of metallic iron
needles.\footnote{%
 However, the survival of metallic iron grains in 
 the ISM is questionable. According to Jones (1990),
 in the ISM which is comparatively rich in atomic oxygen,
 iron grains will undergo rapid oxidation to iron(II)oxide (FeO) 
 and subsequently magnetite (Fe$_3$O$_4$) 
  and even haematite (Fe$_2$O$_3$)
  on time-scales of the order of $10^{6}\yr$;
  the reaction of metallic iron particles with
  sulphur in the ISM will lead to their rapid 
  degradation to sulphide and even sulphate.
 }

\begin{figure}[ht]
\begin{center}
\epsfig{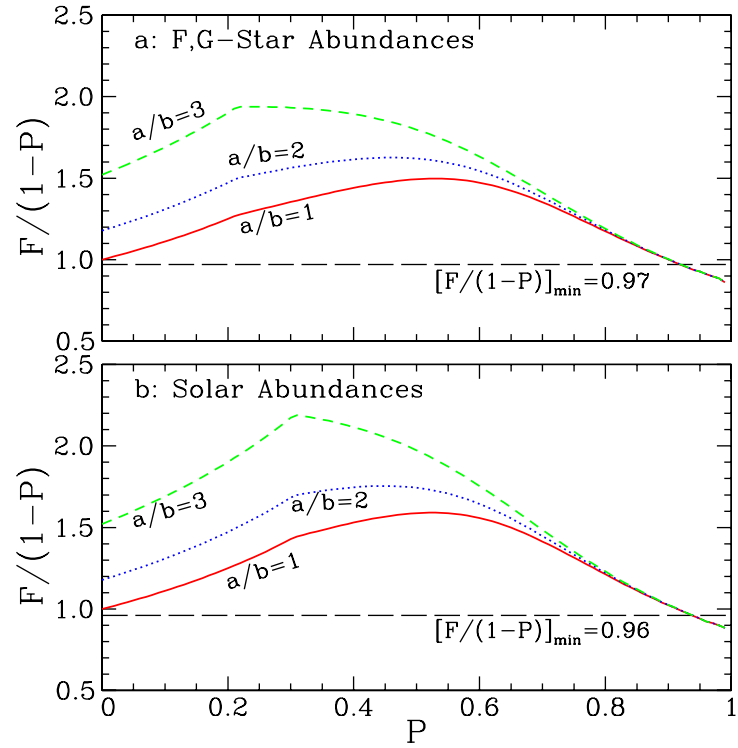}
\end{center}\vspace*{-1em}
\caption{
        \label{fig:FGSun}
        \footnotesize
        Same as Fig.\,\ref{fig:FP}a but for
        models (a) with the F and G star abundances
        or (b) with the solar abundances as the interstellar
        reference standard.
        }
\end{figure}

We have also considered the F and G star abundances
or the Holweger (2001) solar abundances 
(see Table \ref{tab:sunabd})
as the interstellar reference standard. 
We take all the dust-phase Mg, Si, and Fe atoms 
to be in amorphous silicates and assign four O atoms for 
the average of the Mg, Si, Fe abundances (based on olivine). 
We take all the C atoms remaining from the gas to form 
graphite. The Kramers-Kronig relations place a lower bound 
of $\left[F/\left(1-P\right)\right]_{\rm min}\approx 0.97$ 
for models using the F and G star abundances 
and $\left[F/\left(1-P\right)\right]_{\rm min}\approx 0.96$ 
for models using the solar abundances. 
As shown in Figure \ref{fig:FGSun}, the model-predicted
values of $F/\left(1-P\right)$ always exceed 
the lower limit $\left[F/\left(1-P\right)\right]_{\rm min}$
except for very porous dust with $P>0.92$,
indicating that the F and G star abundances or
the solar abundances are not inconsistent with 
the interstellar extinction.

If we do not adopt the Holweger (2001) 
solar abundances but the most recently 
determined solar values of C 
(Allende Prieto et al.\ 2002)
and O (Asplund et al.\ 2004), 
and assume that the solar abundances 
of other elements like Mg, Si and Fe 
are also close to those of B stars 
like C and O, the solar abundances 
would also fall short in accounting for 
the interstellar extinction.\footnote{%
  The latest solar abundances compiled
  by Asplund, Grevesse, \& Sauval (2005)   
  have C and O close to those B stars,
  but Mg, Si and Fe close to those of
  Holweger (2001). See Table \ref{tab:sunabd}.
  If we adopt those of Asplund et al.\ (2005),
  the interstellar extinction can be
  accounted for by modestly elongated 
  ($a/b\simali$2--3) grains with a vacuum volmue 
  fraction of $\simali$40--60\%.
  }
Given that even the elemental abundances 
of our own Sun are still subject to major 
systematic uncertainties, we are not at 
a position to conclude that the solar or 
F and G star abundances are a more preferable
reference standard for the ISM. 

It is very possible that the abundances 
of refractory elements in stars and in 
the ISM are different, namely, there does not
exist any set of abundances of refractory elements 
derived from {\it stellar atmospheres}, 
minus the measured gas-phase abundances of these elements 
in the diffuse ISM, can yield the number of dust atoms 
implied by interstellar extinction.
Various physical processes such as radiation pressure, 
ambipolar diffusion, and gravitational sedimentation 
occurring during the early stages of star formation 
(Draine 2004; Snow 2000) can act to prevent grains
from being incorporated into stars and, therefore, 
lead to lower abundances of heavy elements in stars 
than in the ISM out of which they have formed. 
Alternatively, the stellar {\it photospheric}
abundances of heavy elements may under-represent 
the composition of the star, due to the possible
underestimation of the degree of heavy-element settling 
in stellar atmospheres.
The latter is also indicated by the inconsistency
between the most recent solar abundance determinations
and the helioseismological results (Bahcall et al.\ 2004).

However, there does exist some evidence for the subsolar nature
of the interstellar abundances: 
some noble gases have subsolar abundances in the ISM
(e.g. Cardelli \& Meyer [1997] found that krypton has 
an interstellar abundance of 60\% of solar; 
neon was found to be $\simali$75\% solar 
in the ISM [Takei et al.\ 2002]).   
Since noble gases are not likely to be depleted onto dust,
the abundances of noble gases in the ISM should reflect 
the total interstellar abundances.
The problem of what might be the most appropriate set of 
interstellar reference abundances is still awaiting a solution,
but it is unlikely for the ISM to have abundances as low
as those of B stars since otherwise there will not be enough
material to make the dust needed to explain the observed
interstellar extinction. 

\acknowledgments
I thank G.J. Bendo, J.I. Lunine, K.A. Misselt,
and the anonymous referee for helpful discussions
and/or suggestions.
I thank the University of Arizona for the ``Arizona 
Prize Postdoctoral Fellowship in Theoretical Astrophysics''.

\clearpage

\clearpage

\clearpage

\begin{table}[h]
\begin{center}
\caption[]{Solar and stellar abundances for the dust-forming
elements C, O, Mg, Si and Fe 
(relative to $10^6$ hydrogen atoms)\label{tab:sunabd}.}
\begin{tabular}{lcccccl}
\hline
\hline
Object & C & O & Mg & Si & Fe & References \\
\hline
Sun	&370
	&676
	&34
	&32
	&34
	&Cameron 1973\\
Sun	&417
	&692
	&39.9
	&37.6
	&38.8
	&Cameron 1982\\
Sun	&363
	&853
	&38.5
	&35.8
	&32.3
	&Anders \& Grevesse 1989\\
Sun	&331
	&676
	&38
	&35.5
	&31.6
	&Grevesse \& Sauval 1998\\
Sun	&391
	&545
	&34.5
	&34.4
	&28.1
	&Holweger 2001\\
Sun	&245
	&490
	&-
	&-
	&-
	&Allende Prieto et al.\ 2002\\
Sun	&-
	&457
	&-
	&-
	&-
	&Asplund et al.\ 2004\\
Sun	&245
	&457
	&33.9
	&32.3
	&28.2
	&Asplund et al.\ 2005\\
\hline
F,G stars &358
	  &445
	  &42.7
	  &39.9
	  &27.9
	  &Sofia \& Meyer 2001\\
\hline
B stars &190
	&350
	&23
	&18.8
	&28.5
	&Sofia \& Meyer 2001\\
\hline
\end{tabular}
\end{center}
\end{table}

\clearpage


\begin{thebibliography}{}
\bibitem[]{773}Allende Prieto, C., Lambert, D.L.,
               \& Asplund, M. 2002, ApJ, 573, L137
\bibitem[]{773}Anders, E., \& Grevesse, N. 1989, 
               Geochim. Cosmochim. Acta, 53, 197
\bibitem[]{}Asplund, M., Grevesse, N., \& Sauval, A.J. 
               2005, in Cosmic Abundances as Records of 
               Stellar Evolution and Nucleosynthesis, 
               ed. F.N. Bash, \& T.G Barnes
               (San Francisco: ASP), in press
\bibitem[]{720}Asplund, M., Grevesse, N., Sauval, A.J., 
            Allende Prieto, C., \& Kiselman, D. 
            2004, A\&A, 417, 751
\bibitem[]{}Bahcall, J.N., Basu, S., Pinsonneault, M., 
            \& Serenelli, A.M. 2004, ApJ, in press 
\bibitem[]{775}Bohren, C.F., \& Huffman, D.R. 1983, Absorption and 
            Scattering of Light by Small Particles (New York: Wiley)
\bibitem[]{777}Cameron, A.G.W. 1973, Space Sci. Rev., 15, 121
\bibitem[]{778}Cameron, A.G.W. 1982, in Essays in Nuclear
            Astrophysics, ed. C.A. Barnes, D.D. Clayton, 
            \& D.N. Schramm (Cambridge: Cambridge Univ. Press), 23 
\bibitem[]{781}Cardelli, J.A., \& Meyer, D.M. 1997, ApJ, 477, L57
\bibitem[]{782}Draine, B.T. 2003a, in The Cold Universe,
            Saas-Fee Advanced Course Vol.\,32, 
            ed. D. Pfenniger (Berlin: Springer-Verlag), 213
\bibitem[]{785}Draine, B.T. 2003b, ARA\&A, 41, 241
\bibitem[]{785}Draine, B.T. 2004, in ASP Conf. Ser. 309,
            Astrophysics of Dust, ed. A.N. Witt, 
            G.C. Clayton, \& B.T. Draine 
            (San Francisco: ASP), 691
\bibitem[]{786}Draine, B.T., \& Lee, H.M. 1984, ApJ, 285, 89
\bibitem[]{787}Dwek, E. 1997, ApJ, 484, 779
\bibitem[]{788}Greenberg, J.M. 1974, ApJ, 189, L81
\bibitem[]{789}Greenberg, J.M., \& Li, A. 1996, A\&A, 309, 258
\bibitem[]{790}Grevesse, N., \& Sauval, A.J. 1998, 
            Space Sci. Rev., 85, 161
\bibitem[]{792}Hildebrand, R.H., \& Dragovan, M. 1995, ApJ, 450, 663
\bibitem[]{793}Holweger, H. 2001, in Solar and Galactic
            Composition, ed. R.F. Wimmer-Schweingruber 
            (Berlin: Springer), 23 
\bibitem[]{796}Jones, A.P. 1988, MNRAS, 234, 209
\bibitem[]{796}Jones, A.P. 1990, MNRAS, 245, 331
\bibitem[]{797}Kim, S.-H., \& Martin, P.G. 1996, ApJ, 462, 296
\bibitem[]{798}Lee, H.M., \& Draine, B.T. 1985, ApJ, 290, 211
\bibitem[]{800}Li, A. 2004a, in Penetrating Bars Through Masks 
            of Cosmic Dust: the Hubble Tuning Fork Strikes a New Note, 
            ed. D.L. Block, I. Puerari, K.C. Freeman,
            R. Groess, \& E.K. Block (Dordrecht: Kluwer), 535
\bibitem[]{803}Li, A. 2004b, in ASP Conf. Ser. 309,
            Astrophysics of Dust, ed. A.N. Witt, 
            G.C. Clayton, \& B.T. Draine 
            (San Francisco: ASP), 417
\bibitem[]{807}Li, A., \& Draine, B.T. 2001, ApJ, 554, 778
\bibitem[]{808}Li, A., \& Greenberg, J.M. 1997, A\&A, 323, 566
\bibitem[]{812}Mathis, J.S. 1996, ApJ, 472, 643
\bibitem[]{813}Mathis, J.S. 1998, ApJ, 497, 824
\bibitem[]{814}Mathis, J.S., \& Whiffen, G. 1989, ApJ, 341, 808
\bibitem[]{814}Morton, D.C., Drake, J.F., Jenkins, E.B., 
               Rogerson, J.B., Spitzer, L., \& York, D.G. 
               1973, ApJ, 181, L103
\bibitem[]{815}Ossenkopf, V. 1991, A\&A, 251, 210
\bibitem[]{816}Purcell, E.M. 1969, ApJ, 158, 433
\bibitem[]{817}Roessler, D.M., \& Huffman, D.R. 1998,
            in Handbook of Optical Constants of Solids II,
            ed. E.D. Palik (Boston: Academic), 919 
\bibitem[]{820}Rouleau, F., \& Martin, P.G. 1991, ApJ, 377, 526
\bibitem[]{}Snow, T.P. 2000, J. Geophys. Res., 105, 10239
\bibitem[]{821}Snow, T.P., \& Witt, A.N. 1995, Science, 270, 1455
\bibitem[]{822}Snow, T.P., \& Witt, A.N. 1996, ApJ, 468, L65
\bibitem[]{823}Sofia, U.J. 2004, in ASP Conf. Ser. 309,
            Astrophysics of Dust, ed. A.N. Witt, 
            G.C. Clayton, \& B.T. Draine 
            (San Francisco: ASP), 393
\bibitem[]{827}Sofia, U.J., \& Meyer, D.M. 2001, ApJ, 554, L221
\bibitem[]{828}Sofia, U.J., Cardelli, J.A., \& Savage, B.D. 
            1994, ApJ, 430, 650 
\bibitem[]{830}Steyer, T.R. 1974, PhD thesis, Univ. Arizona
\bibitem[]{831}Takei, Y., Fujimoto, R., Mitsuda, K., \& Onaka, T. 
            2002, ApJ, 581, 307
\bibitem[]{833}Weingartner, J.C., \& Draine, B.T. 2001, ApJ, 548, 296
\bibitem[]{834}Whittet, D.C.B. 1984, MNRAS, 210, 479
\bibitem[]{835}Zubko, V.G., Dwek, E., \& Arendt, R.G. 
          2004, ApJS, 152, 211
\end{thebibliography}
\end{document}